\documentclass{iau} 
\usepackage{graphicx}

\title[The SKA EoR/CD Experiment] %% give here short title %%
{The Square Kilometre Array Epoch of Reionisation and Cosmic Dawn Experiment}

\author[Cathryn M. Trott]   %% give here short author list %%
{Cathryn M. Trott$^{1,2,3}$
}

\affiliation{$^1$International Centre for Radio Astronomy Research, Curtin University, Bentley Australia \\[\affilskip]
$^2$ARC Centre of Excellence for All-Sky Astrophysics (CAASTRO), Redfern Australia \\[\affilskip]
$^3$ARC Centre of Excellence for All Sky Astrophysics in 3 Dimensions (ASTRO 3D), Bentley Australia \\ email: {\tt cathryn.trott@curtin.edu.au} }

\pubyear{2018}
\volume{333}  %% insert here IAU Symposium No.
\jname{Peering towards Cosmic Dawn}
\editors{Vibor Jeli\'c \& Thijs van der Hulst, eds.}
\begin{document}

\maketitle

\begin{abstract}
The Square Kilometre Array (SKA) Epoch of Reionisation and Cosmic Dawn (EoR/CD) experiments aim to explore the growth of structure and production of ionising radiation in the first billion years of the Universe. Here I describe the experiments planned for the future low-frequency components of the Observatory, and work underway to define, design and execute these programs.
\end{abstract}

\firstsection % if your document starts with a section,
\section{Introduction}

The SKA, comprised of a low-frequency interferometer to be built in the Western Australian desert (SKA-Low) and a mid-frequency dish-based interferometer to be built in South Africa's Karoo region (SKA-Mid), endeavours to be the largest general-purpose radio observatory in the world. Spanning frequencies of 50~MHz--20~GHz, the science programs under current planning include all range of physical spatial scales, from study of the polarised emission from exoplanets in our Solar neighbourhood, to neutron stars, Galactic neutral hydrogen, radio galaxies, the intergalactic medium (IGM), the high-redshift Universe, and the overall Cosmology of the Universe (SKA Science Book 2015). Study of the first billion years of the Universe, encompassing the early emission of radiation in the Cosmic Dawn ($z \gtrsim 20$), and the transformation from the neutral to ionised IGM with the ubiquitous emission of ionising photons in the Epoch of Reionisation ($z \sim 20-5$), is a key science driver for SKA-Low. While the current generation of low-frequency interferometers studying this epoch aim to detect the neutral hydrogen emission, the SKA has the capability to explore this era, providing the real astrophysical and cosmological understanding required to complete our understanding of the evolution of structure in the early Universe.

An international team of scientists is working to design and execute the EoR/CD experimental programs, data analysis, signal extraction and signal interpretation. As a subset of the EoR/CD Science Working Group (SWG), a group that represents this community to the SKA Office, the EoR/CD Science Team comprises 60 members from 10 countries, with key involvement from the full suite of EoR astrophysicists: theorists, simulation experts, observers, and data analysts. Together, the Science Team has the full capability and expertise required to execute this challenging science program. In this Proceeding, I outline the three primary avenues of experiment for exploring this era, before describing the Science Team structure, and examples of current work toward the experiments.

\section{EoR/CD experiments}
There are three primary observational paths toward the EoR and Cosmic Dawn with SKA-Low \cite[(Koopmans et al. 2015)]{koopmans15}: statistical detection and parameter estimation (e.g., cylindrically-averaged and spherically-averaged power spectra); direct tomographic imaging of neutral regions \cite[(Mellema et al. 2015; Wyithe et al. 2015)]{mellema15,wyithe15}; 21cm Forest of neutral hydrogen clouds along the line-of-sight to background quasars \cite[(Ciardi et al. 2015)]{ciardi15}.
\begin{enumerate}
\item \underline{Power spectrum and associated statistics}:\\
\noindent This avenue expands the power spectrum approaches of current instruments to higher redshift, accessing the Cosmic Dawn signal. Typically, high redshift signal is weak compared with the sky temperature. It also will deliver a comprehensive exploration of the astrophysical parameter space of brightness temperature fluctuations, while current experiments only have sufficient sensitivity for signal detection, and basic and imprecise parameter estimation.

The experiment has a three-tiered approach: shallow/ultra wide, medium/wide, and deep/pointed. The experiment will be staged to undertake the shallow survey first, yielding early input into improvements in the global sky model, and specific fields for deeper integrations. There will also be capacity to dynamically schedule between the tiers depending on ionospheric and telescope conditions. Figure \ref{fig1} describes the tiered structure, while Table \ref{table:tiers} describes the proposed observational strategy.
\begin{figure}[b]
% \vspace*{-2.0 cm}
\begin{center}
 \includegraphics[width=4.8in]{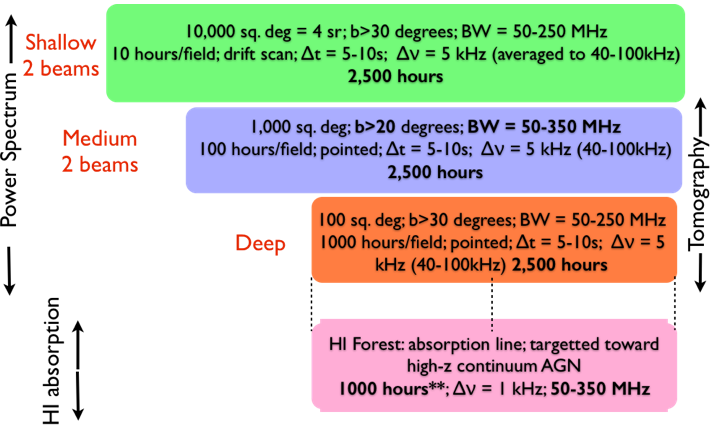} 
 \caption{Proposed tiered structure of the EoR/CD experiments with the SKA. Power spectral analysis is performed over the full bandwidth, while tomography and targeted 21cm-Forest experiments will use the deep pointings.}
   \label{fig1}
\end{center}
\end{figure}
\begin{table}
\centering
\begin{tabular}{|c||c|c|c|c|c|c|}
\hline 
Survey & $z$ & Hours/field & Total hours* & Sky area & Bandwidth & Spectral resolution \\ 
 & & (h) & (h) & (deg$^2$) & (MHz) & (kHz) \\
\hline \hline
Shallow & 5.5--27 & 10 & 2,500 & 10,000 & 150 & 100\\ 
\hline 
Medium & 5.5--27 & 100 & 2,500 & 1,000 & 150 & 100\\ 
\hline 
Deep & 5.5--27 & 1,000 & 2,500 & 100 & 150 & 100 (4.5**)\\ 
\hline 
\end{tabular}
\caption{Proposed survey structure for the tiered experiment. *Assumes two simultaneous beams on the sky, with reduced, shared bandwidth. **\textsc{Hi} Forest spectral resolution within same fields.}\label{table:tiers}
\end{table}
With a constant total correlator capacity, two simultaneous sky pointings (beams) can be formed and processed, at the expense of dividing the full 300~MHz bandwidth into two 150~MHz bands. These overlapping 150~MHz bands provide sufficient bandwidth lever arm for data calibration and foreground excision. The actual power spectrum data processing then occurs over smaller subsets of these datasets, optimised to maximise bandwidth while minimising signal bias due to signal evolution with redshift.

\item \underline{\textsc{Hi} tomography}:\\
The tomography experiment aims to directly map the 21~cm brightness temperature distribution with redshift, yielding cubes of data using a spectral subset of the same dataset from which the statistical experiment is performed (targeting $z=5.5-15$). The sky fields identified for this experiment will have low sky temperature and Galactic diffuse emission, while retaining good calibratability with a high density of point source calibrators. The five sky fields for this experiment will likely be non-contiguous.

\item \underline{\textsc{Hi} Forest}:\\
The \textsc{Hi} (21~cm) Forest experiment will detect and measure the \textsc{Hi} column along the line-of-sight to high-redshift continuum emitters (e.g., QSOs). This experiment relies on the identification of background sources, and may be coincident with the deep pointing survey fields with higher spectral resolution (native narrowband SKA-Low resolution of 4.5~kHz). While the statistical and tomographic experiments provide a broad picture of the structure and evolution of the IGM, the \textsc{Hi} Forest yields targeted information about the topology and evolution of individual structures.
\end{enumerate}

\vspace{0.5cm}
The three experiments therefore cover the suite of local-to-cosmic scales of the evolution of the IGM, while providing an underlying broad band, broad angle, dataset from which other science can be performed (continuum studies, transient studies), and to which novel EoR analyses can be applied. \textit{The wide redshift range of the statistical studies, as well as all of the tomographic and \textsc{Hi} Forest analyses, are new avenues of exploration compared with the current generation of EoR experiments.}

\section{Science Team activities}
\begin{figure}[b]
% \vspace*{-2.0 cm}
\begin{center}
 \includegraphics[width=4.8in]{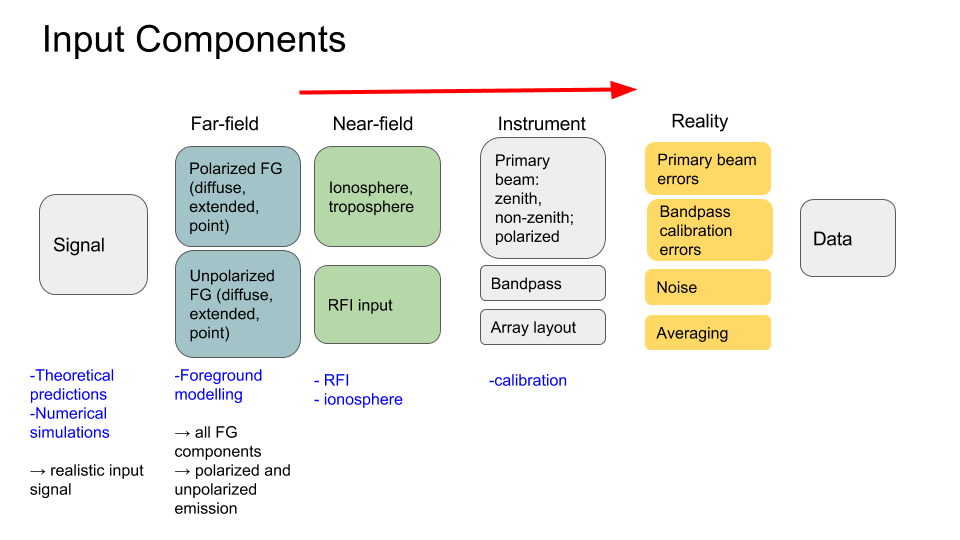}\\
 \includegraphics[width=4.8in]{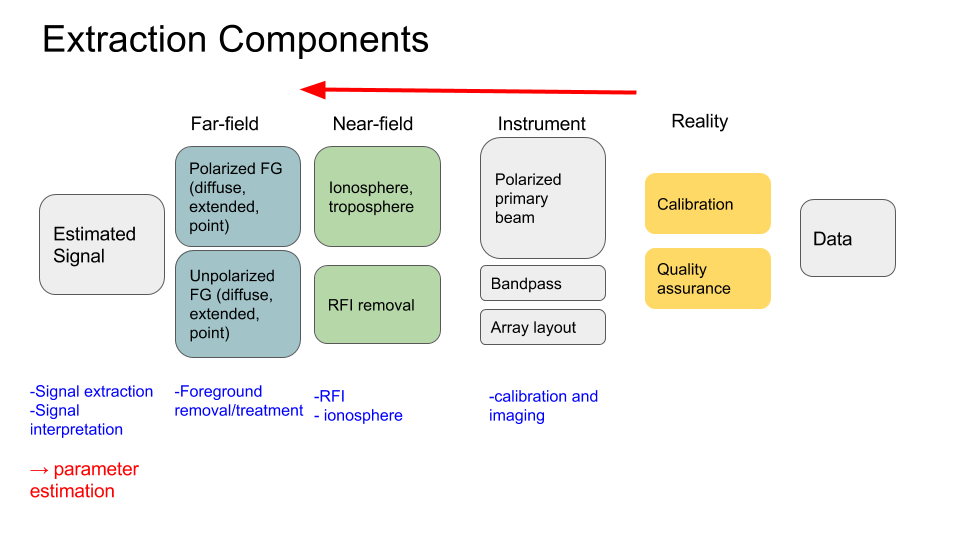}
 \caption{Schematic of Science Team activities, as encompassed by the end-to-end simulation suite under development. (Top) Input components to a realistic set of mock SKA observations (visibilities); (bottom) Output components to extract and interpret the signal.}
   \label{fig:sims}
\end{center}
\end{figure}

The EoR/CD Science Team is actively working toward a Key Science Program (KSP) application for the first five years of science operations. The proposed experiments are as described above, but the group has activity spanning theory, observations, analysis, extraction, and interpretation, underpinning the design and execution of the science experiments. Encompassing all of the activities, an end-to-end simulation suite is being developed to take realistic SKA mock observations of theoretical signals, sky emission, instrument noise and characteristics, process these through existing and under-development pipelines, extract signals, and interpret those in the context of the known inputs. This simulation suite will both test our processing abilities, and identify gaps in the current program of activities. Figure \ref{fig:sims} describes the end-to-end components being actively pursued by the Science Team.

\section{Observational strategy: sky fields}
Augmenting the development of the end-to-end simulation suite, the Science Team is also investing effort in designing the observational strategy for the experiments: e.g., identifying sky fields, defining bandwidths, defining shared observations, setting data quality metrics (ionospheric activity, telescope status) for undertaking different experiments. As part of this program, the Science Team have used existing knowledge of the low-frequency southern radio sky (e.g., through the Murchison Widefield Array GLEAM survey, \cite[Wayth et al. 2015]{wayth15}, and catalogue, \cite[Hurley-Walker et al. 2017]{hurleywalker17}, and diffuse sky models, \cite[Zheng et al. 2017]{zheng17}) to define the optimal observing fields for the tiered experiments.

\begin{figure}[b]
% \vspace*{-2.0 cm}
\begin{center}
 \includegraphics[width=4.8in]{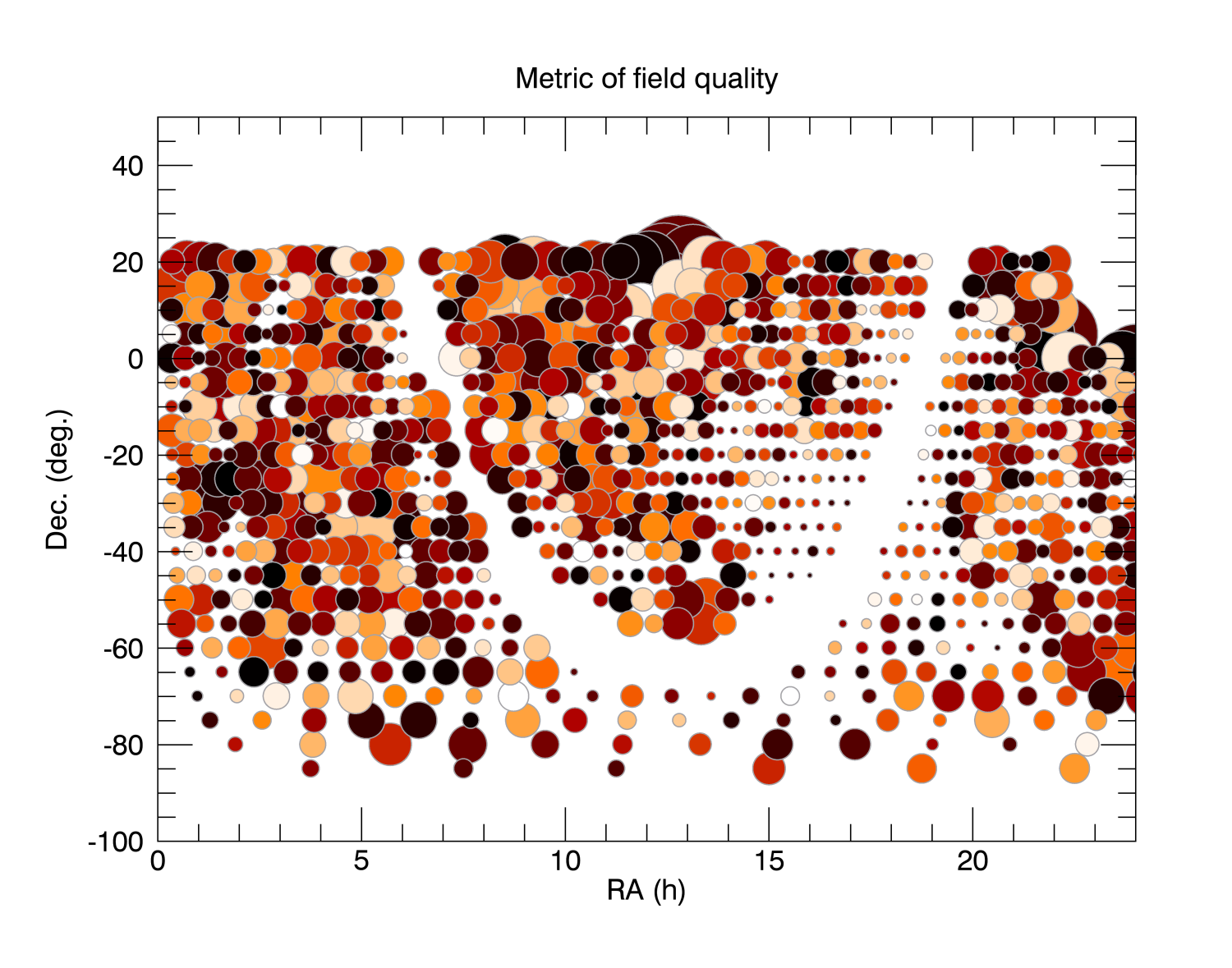} 
 \caption{Metric of observing field quality, combining information about sky temperature, number of extended sources, quality of calibration point sources, and zenith angle.}
   \label{fig2}
\end{center}
\end{figure}

\begin{figure}[b]
% \vspace*{-2.0 cm}
\begin{center}
 \includegraphics[width=4.8in]{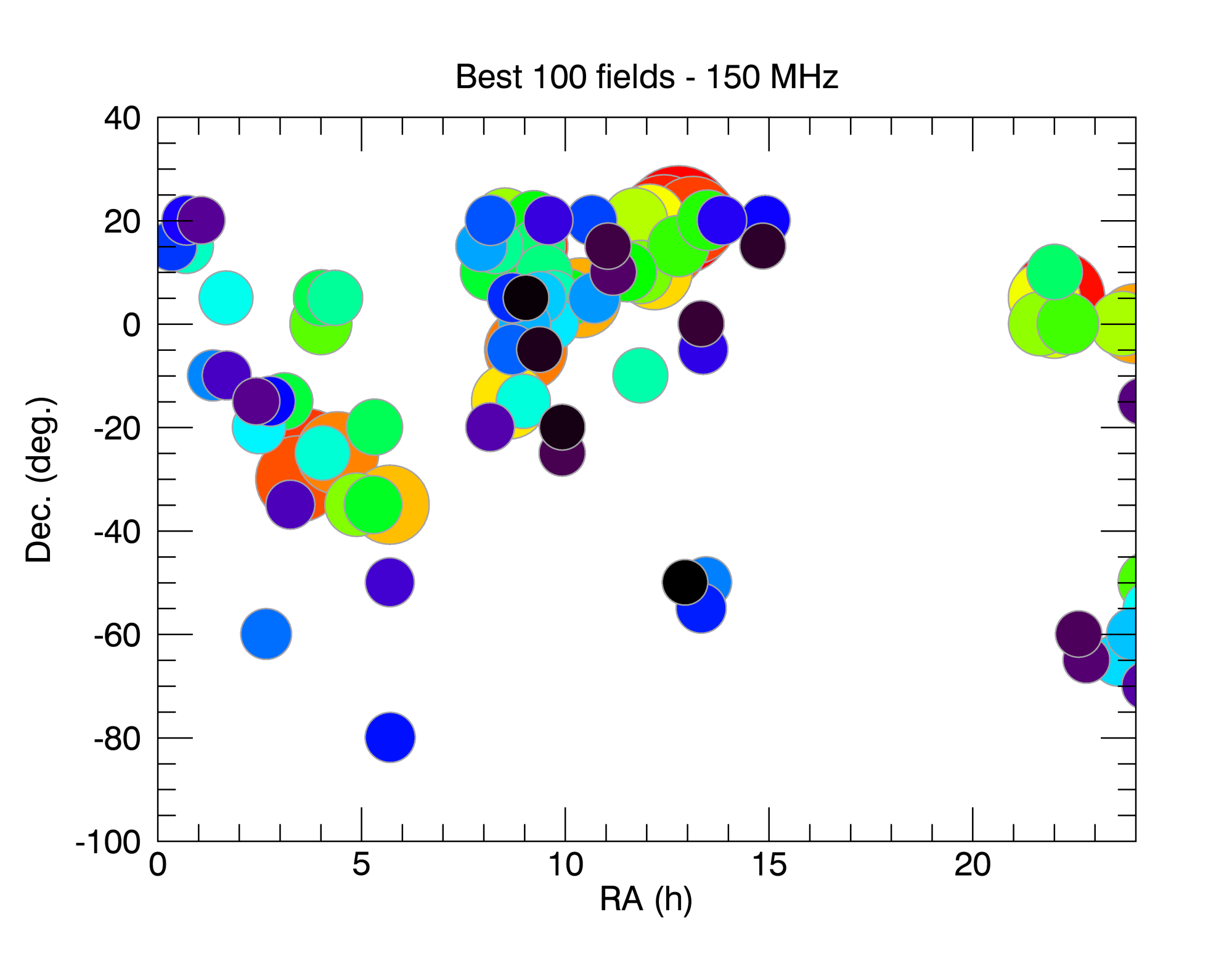} 
 \caption{Best 100 fields for deep EoR/CD integrations, where circle size denotes quality (larger is better). Erroneous measurements are found at the northern-most declination of the GLEAM survey, and large circles at $+$20~degrees should be ignored.}
   \label{fig3}
\end{center}
\end{figure}

A simple metric is employed to obtain a rank-ordered list of potential observing fields, based on the point, extended and diffuse emission in those fields, and down-weighting fields that have low elevation angles. The outputs of this initial assessment are displayed in Figure \ref{fig2} (metric of observing quality, where a larger circle denotes a more favourable field) and \ref{fig3} (best 100 fields at 150~MHz, with the largest circle denoting the most favourable).
Clearly, the Galactic plane should be avoided. There is a concentration of good fields near RA=4~hours, Dec=--30~degrees, consistent with a cold patch of sky that is currently used in the Murchison Widefield Array EoR experiment. Erroneous measurements are found at the northern-most declination of the GLEAM survey, and large circles at $+$20~degrees should be ignored. Note that this work provides only a preliminary assessment of the total intensity fields that could be used for the deep pointings.

\section{Summary and forward look}
The Science Team working within the SKA EoR/CD Science Working Group is proposing a tiered five-year survey, with three major experimental directions: statistical estimation, tomographic imaging, and \textsc{Hi} Forest. The Science Team is actively working to define the observational program, data analysis, signal extraction, and signal interpretation. This suite of work is underpinned by the development of a full end-to-end simulation pipeline to test and refine all components of the experiment. Preliminary investigations are underway to use existing knowledge to define the observational program for the Key Science Experiment.


\begin{thebibliography}{}

\bibitem[Ciardi et al. (2015)]{ciardi15}
{Ciardi, B., Inoue, S., Mack, K., Xu, Y., Bernardi, G.} 2015
\textit{Advancing Astrophysics with the Square Kilometre Array}, AASKA14, 6

\bibitem[Hurley-Walker et~al. 2017]{hurleywalker17}
{Hurley-Walker N.,  et~al.}, 2017, MNRAS, 464, 1146

\bibitem[Koopmans et al. (2015)]{koopmans15}
{Koopmans, L.V.E., et al.} 2015,
\textit{Advancing Astrophysics with the Square Kilometre Array}, AASKA14, 1

\bibitem[Mellema et al. (2015)]{mellema15}
{Mellema, G., Koopmans, L.V.E., Shukla, H., Datta, K.K., Mesinger, A.} 2015
\textit{Advancing Astrophysics with the Square Kilometre Array}, AASKA14, 10

\bibitem[Wayth et~al. (2015)]{wayth15}
{Wayth R.~B.,  et~al.}, 2015, PASA, 32, 25

\bibitem[Wyithe et al. (2015)]{wyithe15}
{Wyithe, J.S.B., Geil, P., Kim, H.} 2015
\textit{Advancing Astrophysics with the Square Kilometre Array}, AASKA14, 15

\bibitem[Zheng et al. 2017]{zheng17}
{Zheng, H., Tegmark, M., Dillon, J. S., Kim, D. A., Liu, A., Neben, A. R., Jonas, J., Reich, P., Reich, W.}, 2017, MNRAS, 464(3), 3486

\end{thebibliography}
\end{document}